# Antimony for broadband nanophotonics across the ultraviolet, visible and infrared

**Fernando Chacón-Sánchez[1,*], Jacek Wojcik[2], Marina García Pardo[1], Fátima Cabello[1], Peter Mascher[2], Fernando Agulló-Rueda[3], Rosalía Serna[1,*], Johann Toudert[1,*]**

*[1] Laser Processing Group, Instituto de Óptica, CSIC, Madrid, Spain*

*[2] Department of Engineering Physics and Centre for Emerging Device Technologies, McMaster University, Hamilton, Ontario, Canada*

*[3]Instituto de Ciencia de Materiales de Madrid (ICMM), CSIC, Madrid, Spain*

**Corresponding authors:*

*fernando.chacon@csic.es, rosalia.serna@csic.es, johann.toudert@csic.es*

**Abstract**. Semimetal elemental antimony (Sb) nanostructures show great potential for applications where nanophotonic properties play a key role, such as phase-change optical memories, non-linear optical elements, photothermal therapy agents, photodetectors and photocatalysts. However, designing advanced Sb-based photonic devices critically requires an accurate and reliable knowledge of the optical response of bulk and nanoscale Sb. Herein, we report for the first time a fully consistent and accurately measured dielectric function for Sb nanoscale films in a wide spectral range from the ultraviolet to the far infrared (4 - 0.04 eV, i.e. ~ 0.3 - 30 μm), surpassing previous reports that explored a limited spectral range. It is found that the Sb spectral response is driven exclusively by giant interband transitions in the visible up to mid infrared (4 - 0.4 eV, i.e. ~ 0.3 - 3 μm), and that their contribution dominates over that of free carriers down to 0.12 eV (i.e. ~ 10 μm). Such spectral response enables Sb nanostructures to display spectrally selective and tunable nanophotonic resonances. First, we showcase interband plasmonic resonances in the visible-to-near infrared for Sb nanogratings. Second, we report giant refractive index dielectric resonances in the mid infrared for nanostructured Sb/dielectric/metal resonant cavities. These findings open a pathway to optimized planar Sb nanoscale designs enabling a tailored light-matter interaction, which will be useful for integrated data, telecom, medical, optoelectronic and energy conversion devices operating in a broad spectral range.

**Keywords:** Antimony, dielectric function, nanophotonics, plasmonics, giant refractive index, interband transitions

## 1. Introduction

Antimony (Sb) compounds were already used at very early stages of History, as cosmetic pigments in Ancient Egypt and later for opacifying glasses during the Hellenistic and Roman periods. Nowadays, they are being used for a broad range of applications, in particular as active media for solar cells, photodetectors and thermoelectric devices. More recently also the interest in pure elemental Sb has grown. In particular, Sb nanostructures including ultrathin ones known as 2D antimony or antimonene [1-4] are considered for applications where nanophotonic properties play a key role, such as phase-change optical memories [5-7], non-linear optical elements [8-11], photothermal therapy agents [12-14], photodetectors [15-16] and photocatalysts [17-18]. However, state-of-the-art performance has not been achieved for such Sb-based devices, in part due to the lack of comprehensive, accurate and reliable information about the optical response of bulk and nanoscale Sb.

Sb is a semi-metal with a rich electronic band structure, which in addition to enabling free carriers that can be probed with far infrared photons, includes several interband transitions that can be triggered by photons at energies from the ultraviolet to far infrared [19-21]. Such interband transitions have the potential to enable a broadband optical absorption suitable for photocarrier generation. Despite these interesting features being reported already in the early 1960s, the optical properties of bulk and nanoscale Sb - in particular its dielectric function $\varepsilon$ - have to date not been accurately and consistently measured across the entire ultraviolet to far infrared range. Previous studies reported $\varepsilon$ in limited spectral regions [5, 21-29], and measuring either not good quality or not well characterized samples (macroscopic crystals for the determination of bulk $\varepsilon$, thin and ultrathin films for the determination of nanoscale $\varepsilon$), thus precluding the use of models suitable for extracting reliable data.

In this work we report *for the first time* the dielectric function $\varepsilon$ of nanoscale Sb across the *whole* ultraviolet-to-far infrared range, which is accurately and reliably measured by spectroscopic ellipsometry and suitable data analysis on high quality continuous, smooth and oriented polycrystalline Sb nanoscale films. We address the underlying physics leading to the observed spectral behaviour of $\varepsilon$ by quantifying the contribution of interband transitions and free carriers. We then analyse *for the first time* its impact on the nanophotonic properties of planar Sb nanostructures: Sb nanogratings and Sb/dielectric/metal nanostructured resonant cavities. We finally offer a perspective on how such properties enable tailored light-matter interaction for optimized Sb-based integrated data, telecom, medical, optoelectronic and energy conversion devices operating in a wide spectral range.

## 2. Results

### 2.1. Sb thin films: surface, structure and optical properties

The studied Sb thin films were grown by pulsed laser deposition (PLD) of Sb in vacuum and at room temperature on a self-passivated (001) silicon substrate. Details on film deposition and characterization are given in the **Methods Section**. **Figure 1a** displays a schematic representation of a selected film. As shown by the AFM image displayed in **Figure 1b**, the film is continuous and its surface roughness is very small, with the peak-to-valley values being smaller than 2 nm.

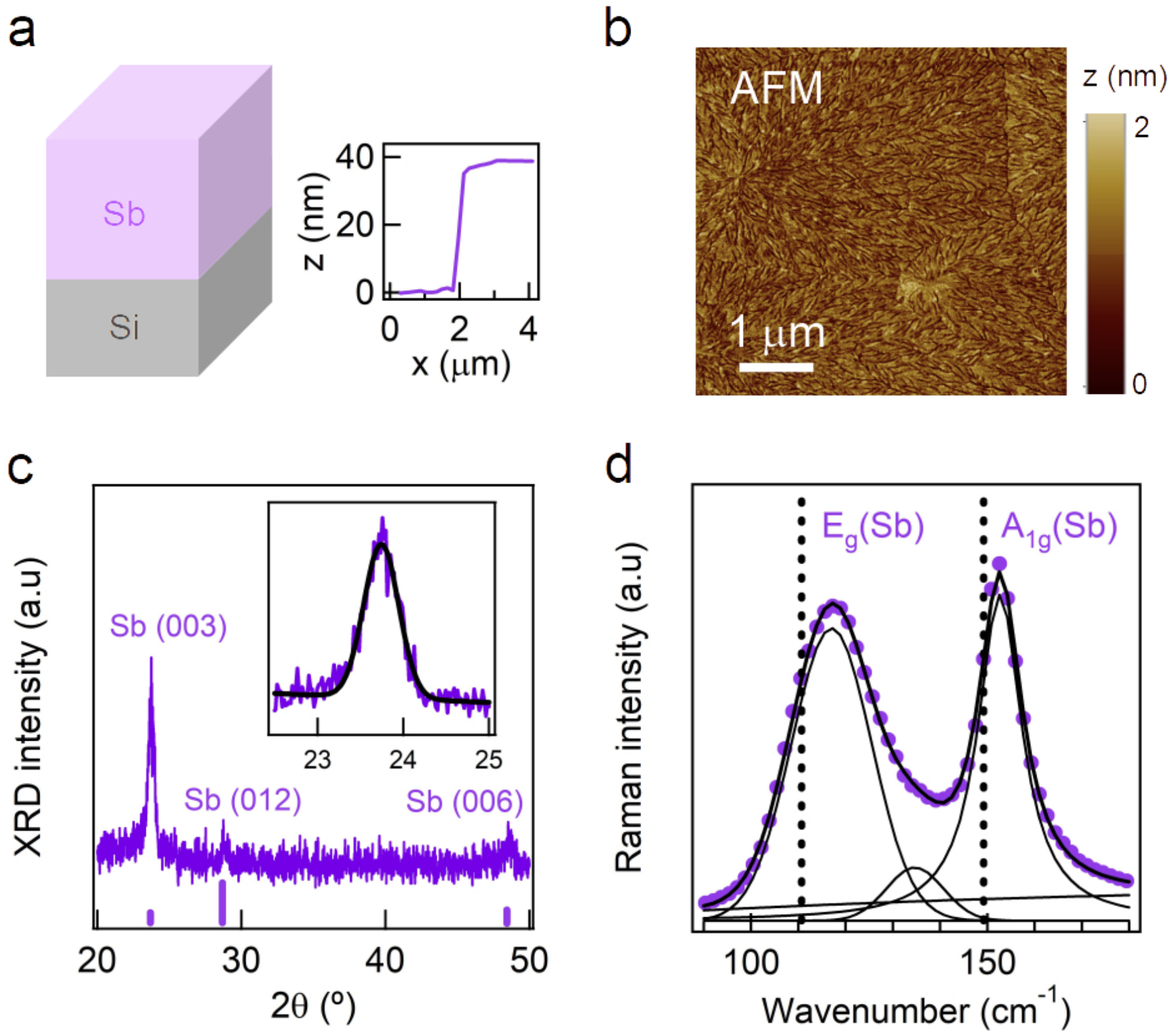

**Figure 1. Structure of a selected Sb thin film.** (a) Schematic representation and step profile measured by atomic force microscopy (b) Atomic force microscopy image of the film's smooth surface. (c) X-ray diffraction pattern of the film. The vertical dashed lines represent the positions of the (003), (012) and (006) powder diffraction peak of Sb taken from JCPDS 00-035-0732. The inset shows the region of the Sb (003) peak. The purple line represents the experimental data. The black line represents the best fit. (d) Raman spectrum of the film in the region of the $E_g$ and $A_{1g}$ modes of Sb. The purple dots represent the experimental data, the thick black line represents the best fit, the thin black lines represent the individual oscillators used for fitting. The vertical dashed lines represent the position of the $E_g$ and $A_{1g}$ modes of a Sb single crystal taken from [30]. The full range Raman spectrum is shown in **Supporting Information S1**.

X-ray diffraction (XRD) suggests that the film consists of crystalline rhombohedral Sb, with mostly a preferential (003) orientation, as demonstrated by the presence of the corresponding peak in **Figure 1c**. Scherrer analysis of the peak width yields a crystallite size of 13.4 nm, showing that the film has a polycrystalline structure. Raman spectroscopy confirms that the film consists of crystalline rhombohedral Sb, as demonstrated by the presence of the $E_g$ and $A_{1g}$ Raman modes of Sb shown in **Figure 1d**. A small additional mode is needed between the two main modes to accurately fit the spectra, which could be attributed to the presence of a minor amount of amorphous Sb. No antimony oxide contribution is seen. Note that the Raman modes appear blue-shifted compared with the corresponding modes of single crystal Sb [29], indicating that the Sb film is under compressive stress, which can be attributed to the high energy of pulsed laser deposited species.

Spectroscopic ellipsometry measurements were performed on the films using two Woollam ellipsometers - a VASE and an IR-VASE - to cover the ultraviolet-to-far infrared region with photon energies ranging from 4 to 0.04 eV. More details are given in the **Methods** section. A realistic three layer (roughness/Sb film/Si substrate) model shown in **Figure 2a** was used to model the ellipsometry data and extract the complex dielectric function $\varepsilon = \varepsilon_1 + i\varepsilon_2$ of Sb using fitting algorithms.

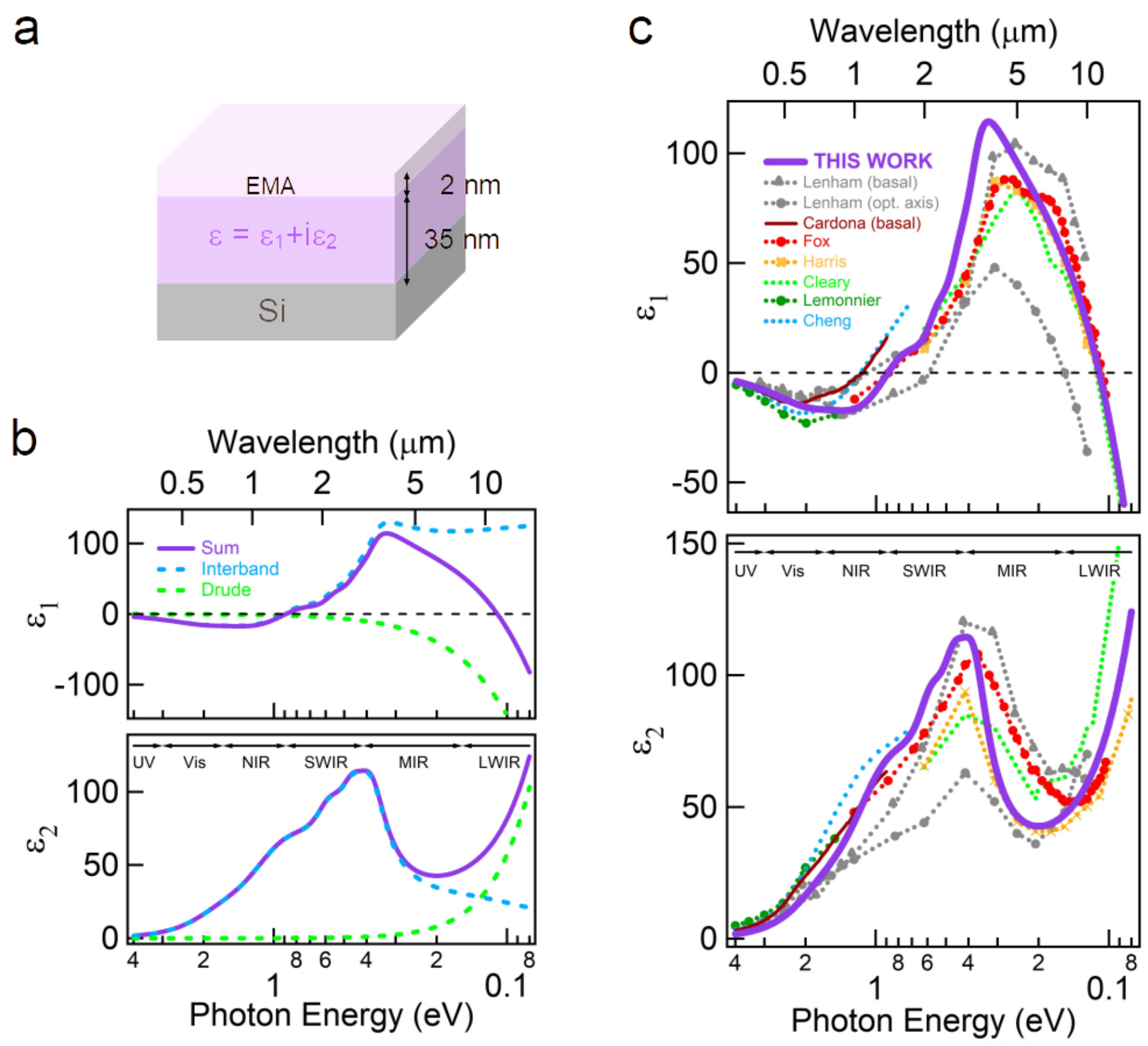

**Figure 2. Optical properties of the selected film measured by spectroscopic ellipsometry from the ultraviolet to the far infrared.** (a) Model used to fit the ellipsometry data. (b) Spectrum of the best fit dielectric function $\varepsilon = \varepsilon_1 + i\varepsilon_2$ («Sum», purple lines), and the contributions of interband transitions («Interband», blue lines) and Drude carriers («Drude», green lines). (c) Comparison between the measured dielectric function of Sb and selected experimental data from the literature by Lenham [22], Cardona [29], Fox [23], Harris [24], Cleary [25], Lemonnier [21], and Cheng [5]. The data reported by Lenham and Cardona were obtained by measurements on cut/polished and freshly cleaved Sb single crystals, respectively. The data reported by Fox were obtained by measurements on cut/polished polycrystals. The data taken from the other works were obtained by measurements on Sb thin films (thickness from few tens of nm to few hundreds of nm).

To enable fitting with few energy-independent parameters, ε was modeled as the sum of 6 Kramers

Kronig-consistent Lorentz oscillators accounting for interband transitions in the spectral range of the measurements, along with a real offset $\varepsilon_\infty$ accounting for interband transitions deeper in the ultraviolet, and 1 Drude oscillator accounting for the contribution of free carriers. The roughness layer was modeled with the Bruggeman effective medium approximation (EMA) mixing 50% of Sb (dielectric function ε) and 50% of air ($\varepsilon_{air} = 1$). In the case of the selected film, the roughness layer thickness was set at 2 nm, according to the peak-to-valley roughness values obtained from **Figure 1**, and the total film thickness was set at 37 nm as determined from step AFM measurements (see **Methods**). An excellent fit of the ellipsometric parameters was achieved, as shown in the **Supporting Information S2**. The values of the fit parameters are provided in **Supporting Information S3**.

The corresponding ε spectrum (plotted in the 4 to 0.08 eV range for a better visualization) is represented in **Figure 2b**, together with the contribution of interband transitions and Drude free carriers, showing that: (i) only the interband contribution plays a role from 4 to 0.4 eV, (ii) both interband and free carrier contributions play a role from 0.4 to 0.12 eV, and (iii) the free carrier contribution dominates below 0.12 eV. Interband transitions result in strong absorption bands that can be mostly seen in the $\varepsilon_2$ spectrum between 0.2 and 1 eV. Because of their high oscillator strength and Kramers Kronig-consistency, they induce negative $\varepsilon_1$ values ($\varepsilon_1 < 0$) from 4 to 0.97 eV, and giant positive $\varepsilon_1$ values ($20 < \varepsilon_1 < 110$) from 0.4 eV to 0.15 eV. In the latter range, however, $\varepsilon_1$ decreases as a result of the free carrier contribution. A faster decrease occurs below 0.12 eV, which turns $\varepsilon_1$ negative again near 0.11 eV. Summarizing, in the 4 to 1 eV range (~ 300 nm to 1.3 μm, ultraviolet-to-near infrared), $\varepsilon_1$ takes negative values that are fully determined by interband transitions. Therefore, Sb displays «interband plasmonic» properties in this spectral region. In the 0.4 to 0.15 eV range (~ 3 μm to 8 μm, mid-infrared), the giant positive values of $\varepsilon_1$ imply that Sb displays a giant refractive index.

The obtained dielectric function ε of Sb is compared in **Figure 2c** with the ones reported in selected previous experimental works [5, 21-25]. This figure illustrates that our dielectric function covers an unprecedently broad spectral region. Furthermore, our work is the first to determine the dielectric function of Sb in a fully accurate and reliable way, because our measurements were done on a well-characterized continuous, smooth and oriented crystalline film, and because the model takes into account the actual film thickness and small roughness, while including Kramers Kronig consistency. This contrasts with previous works (see **Supporting Information S4**), where the material's structure (continuity, crystallinity, roughness, thickness) was not fully characterized thus precluding the use of a suitable model, while Kramers Kronig consistency was not always ensured. As a consequence of this collection of issues, namely unsuitable modeling, lack of structural information, and likely different structure, these previous works reported markedly different values of $\varepsilon_1$ and $\varepsilon_2$ in the spectral ranges where their data overlap.

The historically most relevant of these works was the one reported by Lenham *et al.* [22], who measured the optical reflectance in the spectral range from 3.1 to 0.11 eV of a macroscopic single crystal cleaved along its basal plane and vertical axis, to determine the ordinary and extraordinary bulk optical properties of Sb. While this work is highly valuable because it revealed the main interband transitions of Sb (that peak in the short wave-to-mid infrared) and a strong optical anisotropy, the reported data should be taken with caution. The optical constants reported by Lenham *et al.* were obtained via ellipsometry on mechanically polished single-crystal surfaces. Importantly, polishing was done with a 100 nm finish, leaving a rough and possibly contaminated surface layer, which has a significant effect on the measured reflectance spectrum. However, the surface topography of the crystal was not characterized, and the authors did not take into account the presence of the resulting rough surface layer in their analysis. Therefore, they may have determined a pseudo-dielectric function $\langle\varepsilon\rangle = \langle\varepsilon_1\rangle + i\langle\varepsilon_2\rangle$ that was not solely driven by the bulk dielectric function $\varepsilon$ of Sb but also accounted for the roughness of the surface layer. Similar considerations apply to the work of Fox [23], who did ellipsometry measurements on mechanically polished macroscopic polycrystals. In contrast with the works from Lenham and Fox, Cardona and Greenaway [29] did measurements on freshly cleaved Sb single crystals to obtain the in-plane bulk dielectric function of Sb. This likely enabled them to achieve a smaller surface roughness and hence their analysis included less relevant surface-related artefacts. However, they obtained $\varepsilon$ by Kramers-Kronig transform of a standard reflectance spectrum measured in a range that did not include the main interband transitions of Sb. Therefore, it is very likely that their analysis yielded inaccurate values, especially for $\varepsilon_1$ which strongly depends on the interband transitions located outside of the measured spectral range. In sum, although measurements on macroscopic crystals have provided key qualitative information about the bulk optical properties of Sb, they likely failed in providing quantitatively accurate values of the bulk dielectric function.

The same conclusions can be drawn with respect to measurements on Sb films, found in the works of Harris (reflectance and transmittance, [24]), Cleary (ellipsometry, [25]), Lemonnier (reflectance, [21]) and Cheng (ellipsometry, [5]). None of these works covers the full ultraviolet-to-far infrared range. Cheng focused on the near ultraviolet-visible-near infrared region and thus did not capture the main interband transitions of Sb. These transitions were revealed by Harris and Cleary who focused on the near-to-far infrared region for films in the 20-200 nm and 360-780 nm thickness range, respectively. They peak in the short wave-to-mid infrared as in the case of macroscopic crystals, however with smaller $\varepsilon_2$ values. While Cleary did not report data about crystallinity, Harris mentioned that their films were polycrystalline with a (003) orientation. Importantly, in both works the surface roughness was unknown and hence not taken into consideration in the model. Therefore, the nanoscale $\varepsilon$ values reported in these two works are also likely inaccurate.

In our work, we have prepared Sb films preferentially (003) oriented so that the basal plane lies parallel to the substrate, and therefore by ellipsometry we characterized the in-plane optical response. Interestingly, the contribution of interband transitions reported by Lenham [22] and Harris [24] for the in-plane optical response appears red-shifted with respect to that observed in our work. This difference might result from the fact that the pseudo-dielectric function $\langle\varepsilon\rangle$ of a material with surface roughness displays red-shifted spectral features when compared with the intrinsic $\varepsilon$ of the same smooth material [31]. A similar red-shift is observed when comparing the data from Fox [23] and Cleary [25] with our measurements.

In sum, we have measured for the first time in an accurate, consistent and reliable way the dielectric function of crystalline Sb across the whole ultraviolet-to-far infrared range, in nanoscale films. In the following sections, we analyse for the first time its impact on the nanophotonic properties of planar Sb nanostructures - Sb nanogratings and Sb/dielectric/metal nanostructured resonant cavities.

### 2.2. Sb nanograting: nanophotonic properties in the visible and near infrared

There is scarce information in the literature about the nanophotonic properties of Sb nanostructures in the visible and near infrared. Moschenko *et al.* reported simulations of isolated Sb nanospheres showing resonances in the visible that they attributed to surface plasmons [32]. These authors did not mention what dielectric function they used in their simulations. From simulations using the dielectric functions of Sb from Lemonnier *et al.* [21] and Fox *et al.* [23], our group proposed that these resonances are due to interband plasmons [33]. Huang *et al.* [12] and Chen *et al.* [13] observed and modeled such resonances in Sb nanopolyhedrons dispersed in liquids. In contrast, no information has been reported so far about planar plasmonic Sb nanostructures, despite the relevance of such structures for integrated devices. Interestingly, due to this lack of knowledge, no integrated plasmonic device based in Sb operating in the ultraviolet, visible and near infrared has been reported yet.

Therefore, we assess the potential of planar Sb nanostructures for plasmonics by comparing the visible-to-near infrared dielectric function of Sb with that of key plasmonic materials, as shown in **Figure 3a**. Sb displays negative $\varepsilon_1$ values in the visible-to-near infrared (500 nm to 1300 nm), with $\varepsilon_1$ becoming positive deeper in the infrared. The $\varepsilon_2$ spectrum displays significant values increasing with the wavelength, as a result of interband transitions occurring in the infrared. This spectral behaviour, while different from the one of the reference plasmonic metal, silver (Ag) [34], is comparable to the one of the semimetal bismuth (Bi), the neighbour of Sb in the periodic table and archetypal interband plasmonic material [31, 35]. Bi is successfully used for the design of planar plasmonic nanostructures with broad resonances for light harvesting and subtractive structural color generation, with a performance matching or exceeeding that of noble metals [36-38]. Therefore, thanks to the similarity of its visible-to-near infrared optical response with Bi, Sb is an attractive candidate for the fabrication of planar nanostructures displaying suitable plasmonic properties the visible-to-near infrared.

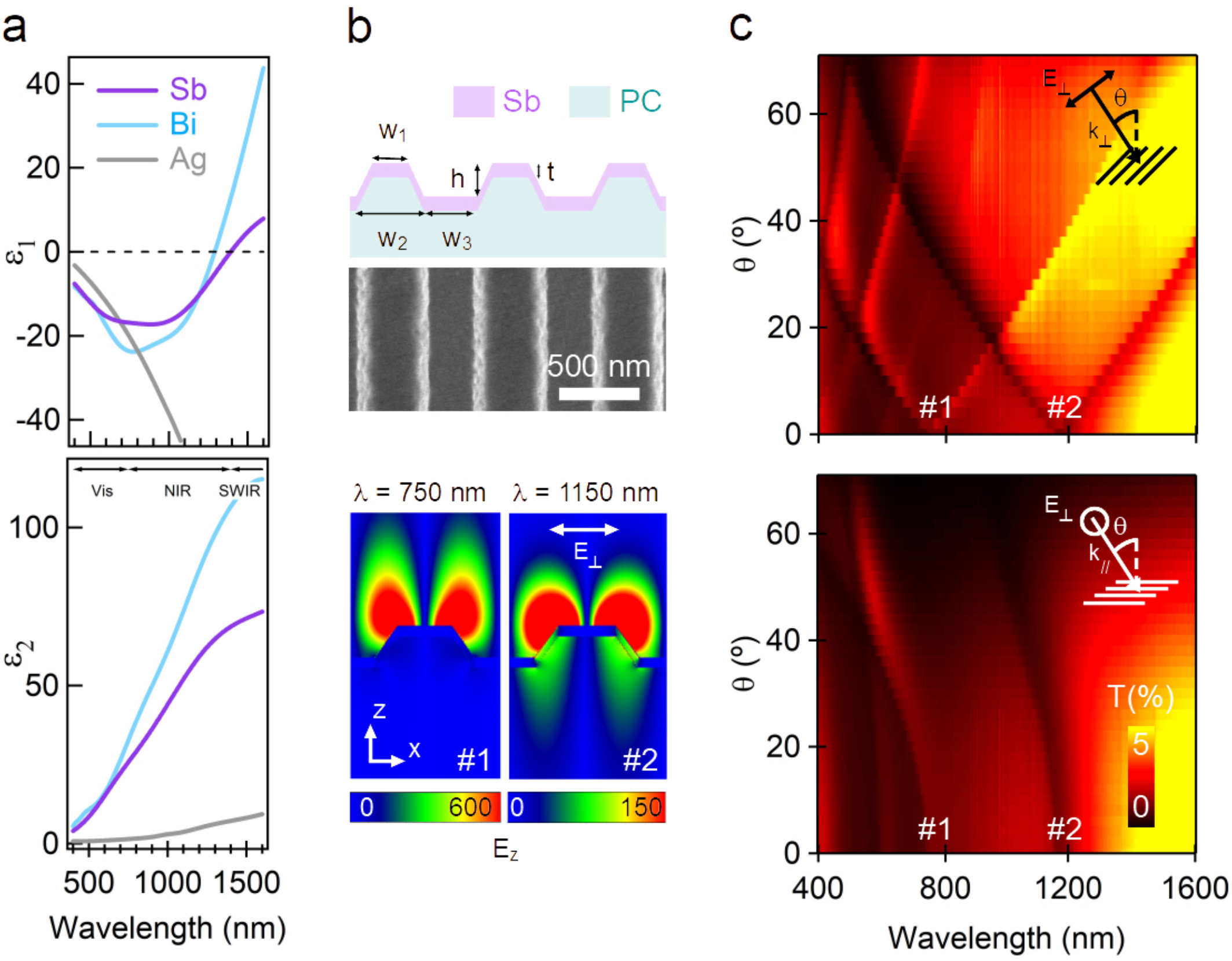

**Figure 3. Sb nanograting: nanophotonic properties in the visible and near infrared.** (a) Spectrum of the dielectric function $\varepsilon = \varepsilon_1 + i\varepsilon_2$ of Sb taken from **Figure 2**, plotted as a function of wavelength across the visible and near infrared. The spectra of Bi and Ag are shown for comparison. These spectra were plotted from the data given in [31] and [34], respectively. (b) Structure and scanning electron microscope image of the Sb nanograting grown on a polycarbonate (PC) template. The grating dimensions determined by atomic force microscopy are $w_1$ = 250 nm, $w_2$ = 455 nm, $w_3$ = 245 nm, h = 165 nm. The Sb film thickness is t = 55 nm as determined on a reference film. These dimensions were used as input to simulate by FDTD the grating's electric near-field patterns at the wavelengths of 750 nm and 1150 nm, at normal incidence for the incident electric field oriented perpendicular to the grating lines ($E_\perp$). These patterns correspond to two plasmonic modes (mode #1 and #2). (c) Color maps of the optical transmittance of the Sb nanograting measured as a function of wavelength and of the angle of incidence θ with the incident electric field oriented perpendicular to the lines ($E_\perp$), for the incident wavevector oriented in the plane perpendicular to the lines ($k_\perp$) or parallel to them ($k_{//}$).

To demonstrate that planar Sb nanostructures display plasmonic properties in the visible-to-near infrared, we studied the structure and nanophotonic response of a simple Sb nanograting structure, grown by conformal pulsed laser deposition of Sb in vacuum and at room temperature on a polycarbonate (PC) template. The periodic grating structure is depicted in **Figure 3b**, where a corresponding scanning electron microscopy (SEM) image is also shown. Finite difference time domain (FDTD) simulations of this structure, using the dielectric function of Sb taken from **Figure 2**, show the electric near-field patterns at the wavelengths of 750 nm and 1150 nm, at normal incidence with the incident electric field oriented perpendicular to the grating lines ($E_{\perp}$). These patterns correspond to two different plasmonic modes (numbered #1 and #2). #1 is a conventional plasmonic mode showing field enhancement at the grating's top interface. In contrast, #2 displays a somewhat less conventional behaviour, as it leads to field enhancement on both the grating's top and bottom interfaces. This behaviour is enabled by the fact that, at 1150 nm, $\varepsilon_1$ nearly cancels out, thus enabling epsilon-near-zero funneling [39], and the subsequent excitation of a plasmon at the bottom interface.

The excitation of these two modes by the $E_{\perp}$ field is confirmed by the experimental transmittance of the nanograting, which is plotted as a function of wavelength and the angle of incidence (θ) in the color maps of **Figure 3c**, and as spectra in **Supporting Information S5**. The modes are seen for $\theta = 0^{o}$ at similar wavelengths as in the FDTD simulations. The good agreement between simulations and experiments is also demonstrated by the transmittance spectra shown in **Supporting Information S6**. As also shown in **Figure 3c**, the modes split and strongly shift across the visible and near infrared upon increasing θ when the beam's wavevector is oriented perpendicular the grating's lines ($k_{\perp}$). In contrast, no splitting is observed and more moderate shifts occur when the beam's wave vector is oriented along the lines ($k_{//}$). These features indicate that plasmons couple thanks to standard diffraction orders for the $k_{\perp}$ orientation, and thanks to conical diffraction orders for the $k_{//}$ orientation [40]. Note, also in **Figure 3c**, that the resonances related with the #2 mode display an angle-of-incidence-independent red-shift with respect to those related with the #1 mode. This suggests that the #2 mode couples with diffraction orders at the Sb/PC interface, the red-shift resulting from the higher refractive index of PC in comparison with air.

Furthermore, as depicted in the color maps of **Supporting Information S7**, no plasmon modes are observed when the incident electric field is oriented parallel to the lines ($E_{//}$), because such a longitudinally oriented field cannot couple with grating plasmons, which are transverse modes.

In sum, these data provide the first demonstration of plasmons on Sb nanogratings in the visible-to-near infrared, and of their interaction with the grating's diffraction orders. Interestingly, such plasmons result in spectrally narrow resonances, thus making them appealing for plasmonic sensing applications. To showcase the potential of Sb nanogratings for sensing, we have immersed them in liquid with different refractive indices (water and ethanol), and observed important changes in their transmittance as a result of the refractive index change both in experiments and FDTD simulations (**Supporting Information S8**). The very good agreement between simulations and experiments opens the way to using the obtained dielectric function of Sb to design structures with a different spectral response, in particular Sb broadband light absorbers for light harvesting and substractive color generation with a performance exceeding that of noble metals.

### 2.3. Sb/dielectric/metal nanostructured resonant cavities: nanophotonic properties in the mid to far infrared

In the literature, no report can be found about the nanophotonic properties of Sb nanostructures (characteristic dimension < 1 μm) in the mid-to-far infrared. In the broader field of Sb structures, only one report by Cleary *et al.* can be found [25]. It showed surface polaritons on a Sb microstructured grating in the long wave infrared region. Here, we thus start exploring the properties of Sb nanostructures in the mid-to-far infrared by studying the response of one of the simplest designs: Sb/dielectric/metal nanostructured resonant cavities.

These cavities are designed following the same concept as the Bi-based nanostructured resonant cavities reported in one of our previous works [41]. The giant n values of Bi ($7 < n < 10$), across the mid-to-far infrared, along with its moderate k values ($2 < k < 3$), enabled the fabrication of very thin cavities (< λ/10 thick) resonating at a wavelength tunable across the 3 to 20 μm range by adjusting the semimetal/dielectric layer thicknesses. On resonance, such cavities behave as perfect absorbers, in which light is fully absorbed by a deeply subwavelength Bi absorbing layer (less than λ/100 thick). Achieving these properties with such small thicknesses is made possible by a unique fractal destructive interference mechanism, which occurs only when n takes giant values. Remarkably, the very small thicknesses involved enable a given cavity to resonate at a single wavelength in the infrared (no multiple resonance orders), thus displaying a strong absorption in a single selected spectral region and a high reflectance elsewhere. This outstanding response is fully suited for infrared light management applications, which previously required devices based on complex metasurface designs. Given the similarities of the Sb and Bi dielectric response discussed in the **Section 2.1**, we can wonder if such impressive features can be achieved with simple cavities based on Sb layers?

To answer this question, we analyze the spectrum of the complex refractive index $N = n + ik$ of Sb in the mid to far infrared and compare it with the one of Bi. To such aim, we have displayed this spectrum in **Figure 4a**, together with that of Bi and also that of the reference high index material: silicon (Si). The real refractive index of Sb displays giant values ($7 < n < 11$) across the mid infrared, and smaller values at longer wavelengths - still remaining higher than that of Si. The extinction coefficient k of Sb takes moderate values (2 < k < 3) across the mid infrared, and much higher values at longer wavelengths. Therefore, the n and k values of Sb are comparable with those of Bi across the mid infrared. In the long wave infrared and far infrared, Sb displays a smaller n and a higher k than Bi. This difference is due to the important free carrier contribution of Sb, which lowers n and increases k in this range, while it does not play a significant role in Bi for which n and k are chiefly driven by the interband contribution. Therefore, Sb/dielectric/metal nanostructured cavities should enable impressive features as those based on Bi in the mid infrared.

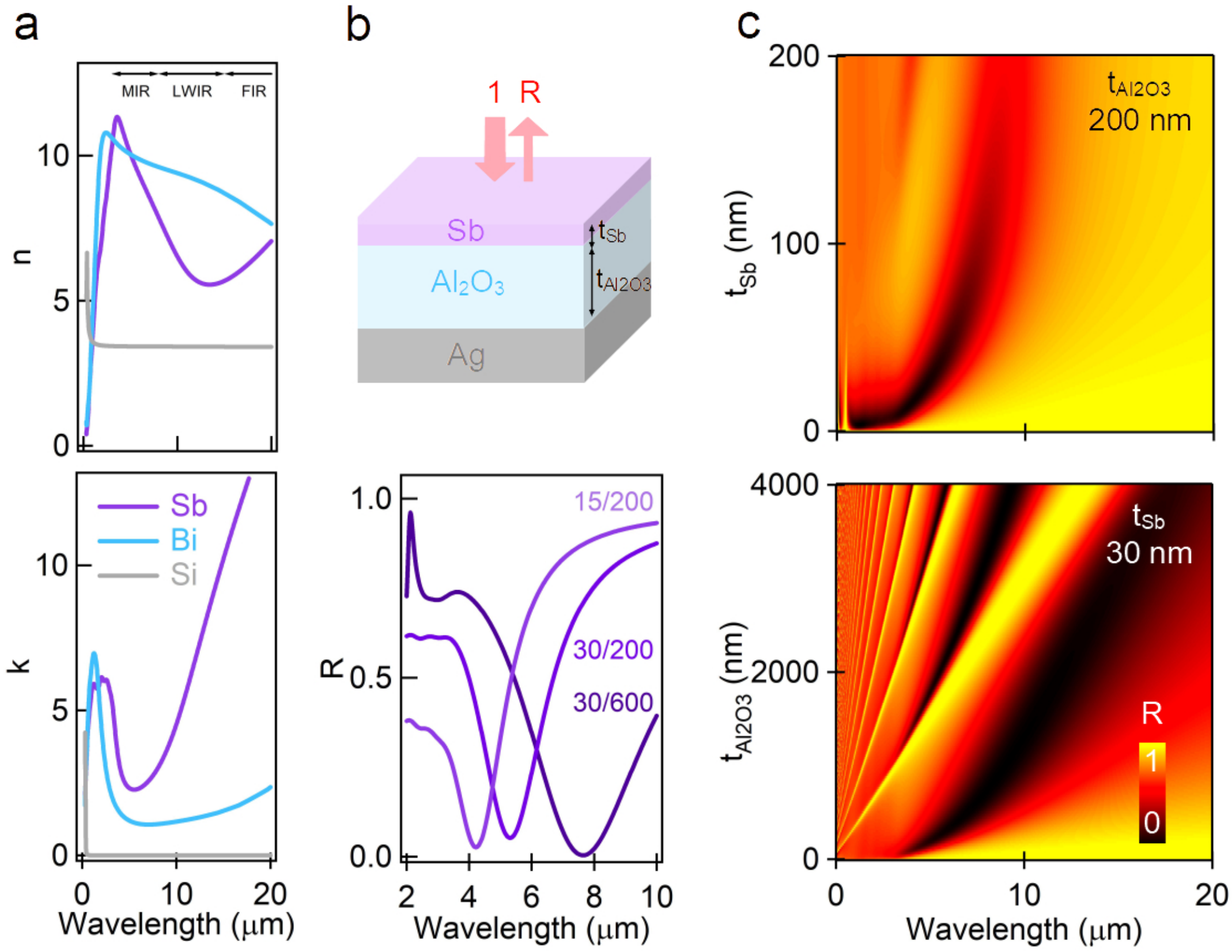

**Figure 4. Sb/dielectric/metal nanostructured resonant cavities: nanophotonic properties in the mid to far infrared.** (a) Spectrum of the complex refractive index N = n + ik derived from the dielectric function of **Figure 2**, plotted as a function of wavelength. The complex refractive index of Bi and Si are shown for comparison. These spectra were plotted from the data given in [40] and [34], respectively. (b) Schematic representation of the considered $Sb/Al_2O_3/Ag$ resonant cavity, and its transfer matrix method (TMM) simulated infrared reflectance spectra at normal incidence for different values of the Sb and $Al_2O_3$ thicknesses ($t_{Sb}/t_{Al2O3}$). The Ag layer is semi-infinite, so that the transmittance T of the cavity is zero, and the absorbance is A = 1 - R. (c) Color maps of the reflectance of the cavity at normal incidence as a function of wavelength and of the Sb and $Al_2O_3$ thicknesses. In the top panel, $t_{Sb}$ was varied while $t_{Al2O3}$ was set at 200 nm. In the bottom panel, $t_{Al2O3}$ was varied while $t_{Sb}$ was set at 30 nm.

To assess this prediction, we have performed transfer matrix method (TMM) simulations of the reflectance spectra of $Sb/Al_2O_3/Ag$ nanostructured resonant cavities at normal incidence. A schematic representation of such cavities is shown in **Figure 4b**. It was assumed that the complex refractive index of Sb does not depend on the Sb layer thickness and is always that shown in **Figure 4a** - an assumption that might fail in the few-nm thickness regime due to quantum effects. The selected reflectance spectra plotted in **Figure 4b** confirm that the cavities display a single resonance resulting in near-perfect absorption (the absorbance A being A = 1 - R, perfect absorption is achieved when R = 0). This resonance is tuned across the mid infrared by varying the Sb and $Al_2O_3$ layer thicknesses ($t_{Sb}$ and $t_{Al2O3}$), with $t_{Sb} < \lambda/100$ and $t_{Sb} + t_{Al2O3} < \lambda/10$. The resonance shift as a function of $t_{Sb}$ and $t_{Al2O3}$ is shown in the color maps of **Figure 4c**. As shown in the top panel, when increasing $t_{Sb}$ from 20 to 200 nm at a fixed $t_{Al2O3}$ (200 nm), the resonance shifts across the mid infrared from short to long wavelengths, with an asymptotic behaviour near the wavelength of 8 μm. As expected, this contrasts with Bi/$Al_2O_3$/Ag nanostructured resonant cavities, in which the resonance shifted linearly with $t_{Bi}$ across the mid-to-far infrared. This difference can be explained by the free carrier induced far infrared absorption occuring in Sb, which hinders the multiple reflections needed for the fractal interference in this range. However, as shown in the bottom panel, this asymptotic behaviour can be bypassed so that the resonance shifts in a broader range, by designing a cavity with a sufficiently thin Sb layer ($t_{Sb}$ = 30 nm) and varying $t_{Al2O3}$. With this approach, however, multiple resonance orders of the cavity appear in the spectrum as $t_{Sb}$ and $t_{Al2O3}$ increase, as its behaviour starts to become similar to that of all-metallic cavities.

In sum, our simulations predict that simple Sb/$Al_2O_3$/Ag nanostructured resonant cavities enable optical resonances with near-perfect absorption tunable across the mid infrared. They are thus appealing for light management in this specific spectral region. This tuning range is smaller than in the case of Bi. However, Sb may take advantage over Bi because of its higher melting point and demonstrated solid-state phase transitions that may be harnessed to achieve active thermal radiation management and infrared switching. These findings lay the foundation for the fabrication and experimental characterization of resonant mid infrared Sb cavities.

**3. Conclusions**

Summarizing, in this work we have measured for the first time in an accurate and reliable way the dielectric function ε of nanoscale Sb from the ultraviolet to the far infrared. The analysis of ε shows that Sb displays giant interband transitions in the near to mid infrared, resulting in interband plasmonic properties in the ultraviolet-to-near infrared, and in a giant refractive index in the mid infrared. For the first time, we have explored the impact of these features on the nanophotonic properties of planar Sb nanostructures: Sb nanogratings and Sb/dielectric/metal nanostructured resonant cavities.

Sb nanogratings display resonant plasmonic modes tunable across the visible-to-near infrared. Although these modes result in spectrally narrow resonances, our analysis suggests that upon suitable design it will be possible to achieve also spectrally broad plasmonic features. Such tunability will enable achieving optimal planar Sb nanostructures for a broad gamut of applications: sensing, phase-change optical memories with enhanced contrast, structural color generation and solar harvesting with a performance exceeeding that of metals.

An even more important feature of Sb is its interband plasmonic character, which is shared by other semiconductors and semimetals [42-45]. It enables the nanostructures to combine a strong and broadly tunable metal-like plasmonic absorption of light with semiconductor-like features, such as an efficient saturable absorption or conversion of the absorbed light into electric energy. Therefore, our findings show a way to design and fabricate optimal planar Sb nanostructures for applications including integrated visible-to-near infrared modulation, photodetection, sunlight-driven photothermal and photocatalytic devices.

Furthermore, thanks to the giant refractive index and moderate extinction coefficient of Sb in the mid infrared, very thin Sb/dielectric/metal nanostructured resonant cavities (Sb thickness $< \lambda/100$, total thickness $< \lambda/10$) display a single resonance tunable by design in this spectral region. This resonance results in near-perfect absorption, while light is near-perfectly reflected off-resonance. This outstanding response is very well-suited for mid infrared light management applications, which previously required devices based on complex metasurface designs. In particular, it can be applied to design optimal planar Sb nanostructures for actively switchable radiative thermal management or thermal sensing.

Therefore, our findings lay down the foundation for the development of planar Sb nanostructure designs harnessing nanophotonic phenomena to achieve a tailored light-matter interaction in a broad spectral range. This opens the way for devices with an optimal performance for various fields such as integrated data, telecom, medical, optoelectronic and energy conversion. In this context, the dielectric function of Sb we provide will serve as a trusted reference for numerical design and a benchmark for material characterization.

**Methods**

The Sb films were grown by pulsed laser deposition on Si substrates. Growth was performed in vacuum at room temperature with a base pressure $P = 3.10^{-7}$ mbar. A high-purity Sb target was ablated with laser pulses generated by a LambdaPhysik LPX 200 excimer laser (wavelength $\lambda$ = 193 nm, pulse duration = 20 ns, frequency f = 5 Hz, energy E = 37.9 mJ). The ablated material was deposited onto single-side polished Si substrates, which were further mechanically roughened at their rough side to fully suppress specular backside reflection during infrared optical measurements.

AFM measurements were performed with a Park Systems XE7 setup. The film thickness was measured by carefully scratching off the film to obtain a clean edge step. For the selected film we measured a thickness of 37 nm.

Raman spectra were obtained in backscattering with a Renishaw Ramascope 2000 spectrometer. The excitation wavelength was 514.5 nm from an argon ion laser. Laser power on the sample was kept below 1 mW to avoid heating. No polarization analyzer was used for the scattered light. The width of the spectrometer slit was 50 μm. The spectra were decomposed in the region of the fundamental phonons with a nonlinear curve fitting to a sum of a cubic background, two Gaussians and a Lorentzian ($A_{1g}$ mode).

Spectroscopic ellipsometry was measured on the Sb thin film with two ellipsometers (J.A. Woollam VASE and IR-VASE) to achieve spectra covering the whole ultraviolet-to-far infrared range (4-0.04 eV). The measurements were done at angles of incidence θ of 55 and 65º. The fitting and analysis were done using the transfer matrix method (TMM) along with Levenberg-Marquardt minimization algorithms integrated in the Woollam WVASE software. As backside reflection was suppressed, the Si substrate was modeled as a semi-infinite medium. The experimental transmittance spectra of the Sb nanograting were measured with the VASE ellipsometer in transmission mode, which enabled controlling both θ and the polarization of the incident light. The nanograting was rotated in its plane to orient the incident beam in a plane parallel or perpendicular to the grating lines. Sensing experiments were performed in a Varian Cary 5000 double beam spectrophotometer equipped with polarizers. The transmittance and near-field maps of the Sb nanograting were modeled by FDTD with the OPTIFDTD software. The reflectance of the Sb/$Al_2O_3$/Ag cavities was simulated using the transfer matrix method with the Woollam WVASE software. All measurements were done at room temperature, and all the dielectric functions used for simulations refer to room conditions.

**Acknowledgements**

This work was supported by the AEI (Spain) under Grant SLIM-2P (PID2024-156974OB-C21) funded by MCIN/AEI/10.13039/501100011033 and by the European Regional Development Fund (ERDF/EU). It was also supported by the Natural Sciences and Engineering Research Council of Canada (NSERC) under its Discovery Grant program (Grant No. RGPIN-2019–06023). It was also supported by the AEI (Spain) under the Severo Ochoa Centers of Excellence Program (CEX2024-001445-S) funded by MCIN/AEI/10.13039/501100011033.

## Supporting Information

# Antimony for broadband nanophotonics across the ultraviolet, visible and infrared

**Fernando Chacón-Sánchez[1,*], Jacek Wojcik[2], Marina García Pardo[1], Fátima Cabello[1], Peter Mascher[2], Fernando Agulló-Rueda[3], Rosalía Serna[1,*], Johann Toudert[1,*]**

*[1] Laser Processing Group, Instituto de Óptica, CSIC, Madrid, Spain*

*[2] Department of Engineering Physics and Centre for Emerging Device Technologies, McMaster University, Hamilton, Ontario, Canada*

*[3]Instituto de Ciencias de Materiales de Madrid (ICMM), CSIC, Madrid, Spain*

**Corresponding authors:*

*fernando.chacon@csic.es, rosalia.serna@csic.es, johann.toudert@csic.es*

## S1. Additional Raman data

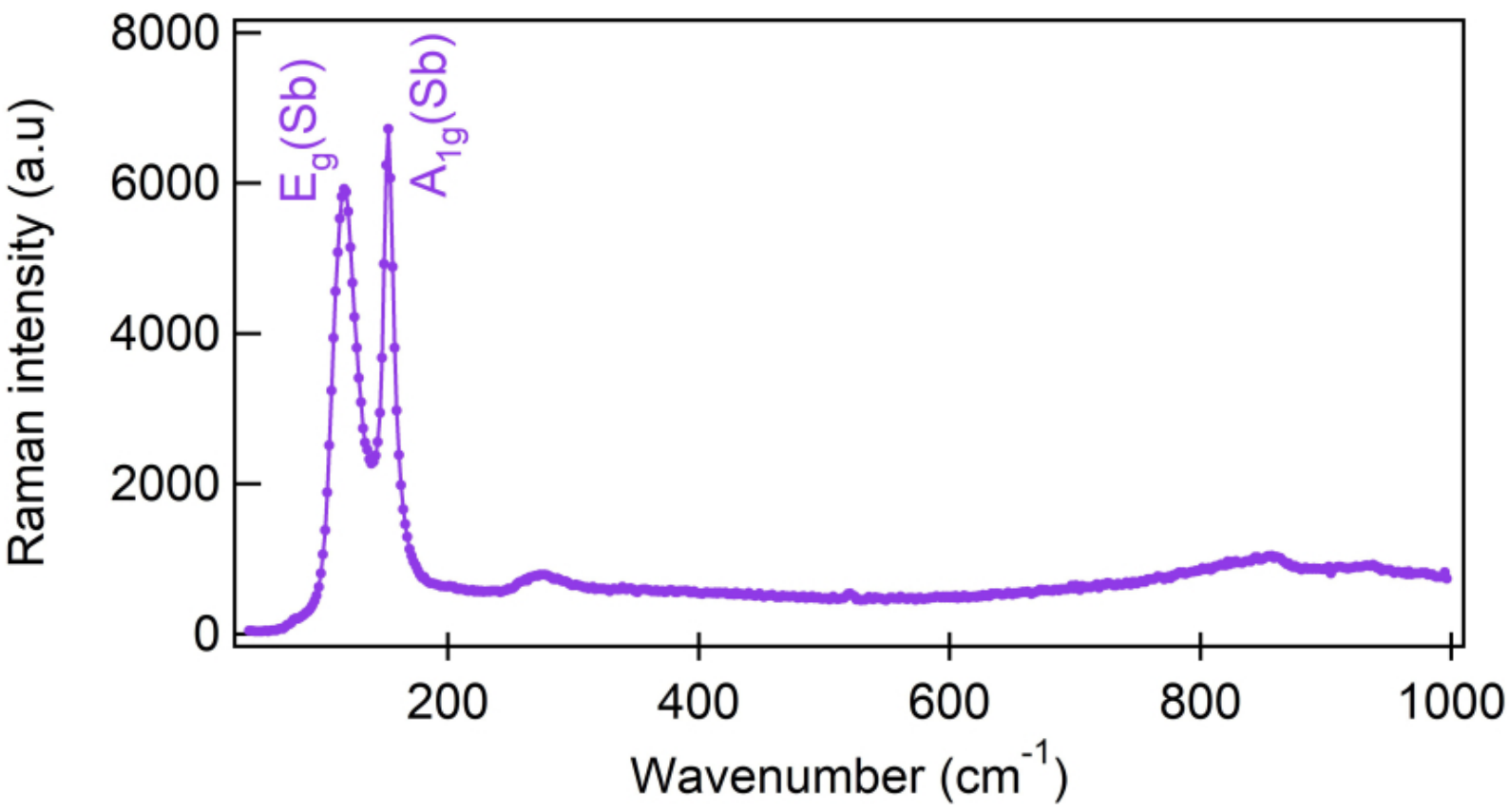

**Figure S1.1. Full measured Raman spectrum of the Sb thin film**

## S2. Spectroscopic ellipsometry spectra

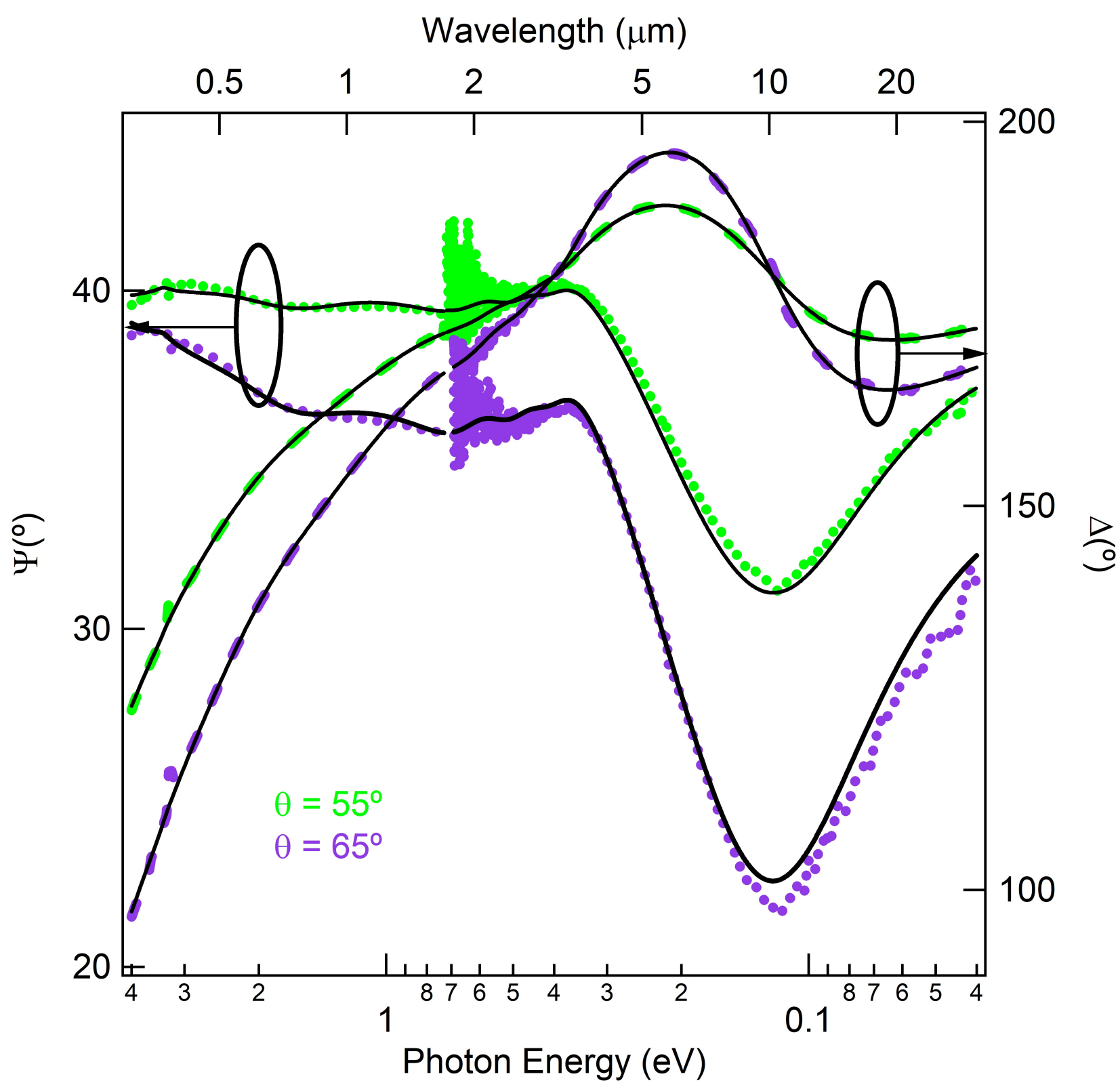

**Figure S2.1. Spectroscopic ellipsometry: measured spectra (dashed green and purple lines) and best fit spectra (black lines)**

**S3. Oscillators used to model the dielectric function of Sb**

$\varepsilon(E) = \varepsilon_\infty + \sum_j \text{Lorentz}_j(E) + \text{Drude}(E)$

With: $\varepsilon_\infty = 1.828$

With: $\text{Lorentz}_j(E) = A_j Br_j En_j / (En_j^2 - E^2 - i Br_j E_j)$

| **Lorentz** | **Amp** | **En (eV)** | **Br** |
|---|---|---|---|
| 1 | 9.315 | 1.793 | 1.455 |
| 2 | 55.45 | 0.965 | 1.042 |
| 3 | 28.98 | 0.585 | 0.224 |
| 4 | 39.09 | 0.463 | 0.164 |
| 5 | 48.86 | 0.383 | 0.134 |
| 6 | 11.45 | 0.174 | 0.295 |

With: $\text{Drude}(E) = -(h/2\pi)^2 / (\varepsilon_0 \rho_n (\tau_n E^2 + i h E / 2\pi))$
Where: $\rho_n = m^*/(N q^2 \tau) = 1/(q \mu N)$

| **Drude** | **N/m* ($cm^{-3}$)** | **$\mu$ ($cm^2/Vs$)** |
|---|---|---|
| 1 | $1.202 \times 10^{21}$ | 29.06 |

## S4. Literature on the optical properties of Sb

| Ref | 1st author | Sample type | Range | Measurement and analysis |
|---|---|---|---|---|
| **THIS WORK** | **Chacón-Sánchez** | **Thin film, pulsed laser deposited, (003) oriented polycrystal, smooth surface** | **UV to FIR 4-0.04 eV** | **Ellipsometry, roughness/film/substrate model, Kramers Kronig-consistent oscillators** |
| [22] | Lenham | Macroscopic single crystal, cut and polished (100 nm finish), unknown roughness | Vis to LWIR 3.1-0.11 eV | Ellipsometry, pseudo-dielectric function, wavelength-by-wavelength inversion |
| [29] | Cardona | Macroscopic single crystal, freshly cleaved | EUV to NIR 25-0.5 eV | Reflectance, pseudo-dielectric function, Kramers-Kronig transform |
| [23] | Fox | Macroscopic polycrystal, unknown orientation and roughness | Near to LWIR 1.2-0.1 eV | Ellipsometry, pseudo-dielectric function, wavelength-by-wavelength inversion |
| [24] | Harris | Thin films (20-200 nm thickness), evaporated, (003) oriented polycrystal, unknown roughness | NIR to FIR 0.61-0.016 eV | Reflectance and transmittance, wavelength-by-wavelength inversion |
| [25] | Cleary | Thin films (340 and 680 nm thickness), evaporated, crystallinity and roughness unknown | NIR to FIR 1.2-0.025 eV | Ellipsometry, pseudo-dielectric function |
| [21] | Lemonnier | Thin film, evaporated, polycristalline, orientation and roughness unknown | EUV to Vis 14.5-2.5 eV | Reflectance, Kramers-Kronig transform |
| [5] | Cheng | Thin and ultrathin films (3-20 nm thickness), evaporated | NUV to NIR 4-0.7 eV | Ellipsometry, film/substrate model, Kramers Kronig-consistent oscillators |
| [26] | Hünermann | Ultrathin films (0.1-2 nm thickness), evaporated, polycrystalline, (003) orientation | MUV to Vis 5.5-1.8 eV | Ellipsometry, pseudo-dielectric function, wavelength-by-wavelength inversion |
| [27] | Resch-Esser | Thin and ultrathin films (0.1-20 nm thickness), evaporated, polycrystalline, (003) orientation | MUV to Vis 5.5-1.8 eV | Ellipsometry, film/substrate model, dielectric function includes roughness contribution, wavelength-by-wavelength inversion |
| [28] | Rossow | Ultrathin films (0.1-8 nm), evaporated, polycrystalline, (003) orientation | MUV to Vis 5.5-1.8 eV | Ellipsometry, film/substrate model, dielectric function includes roughness contribution, wavelength-by-wavelength inversion |

## S5. Sb nanograting: transmittance spectra

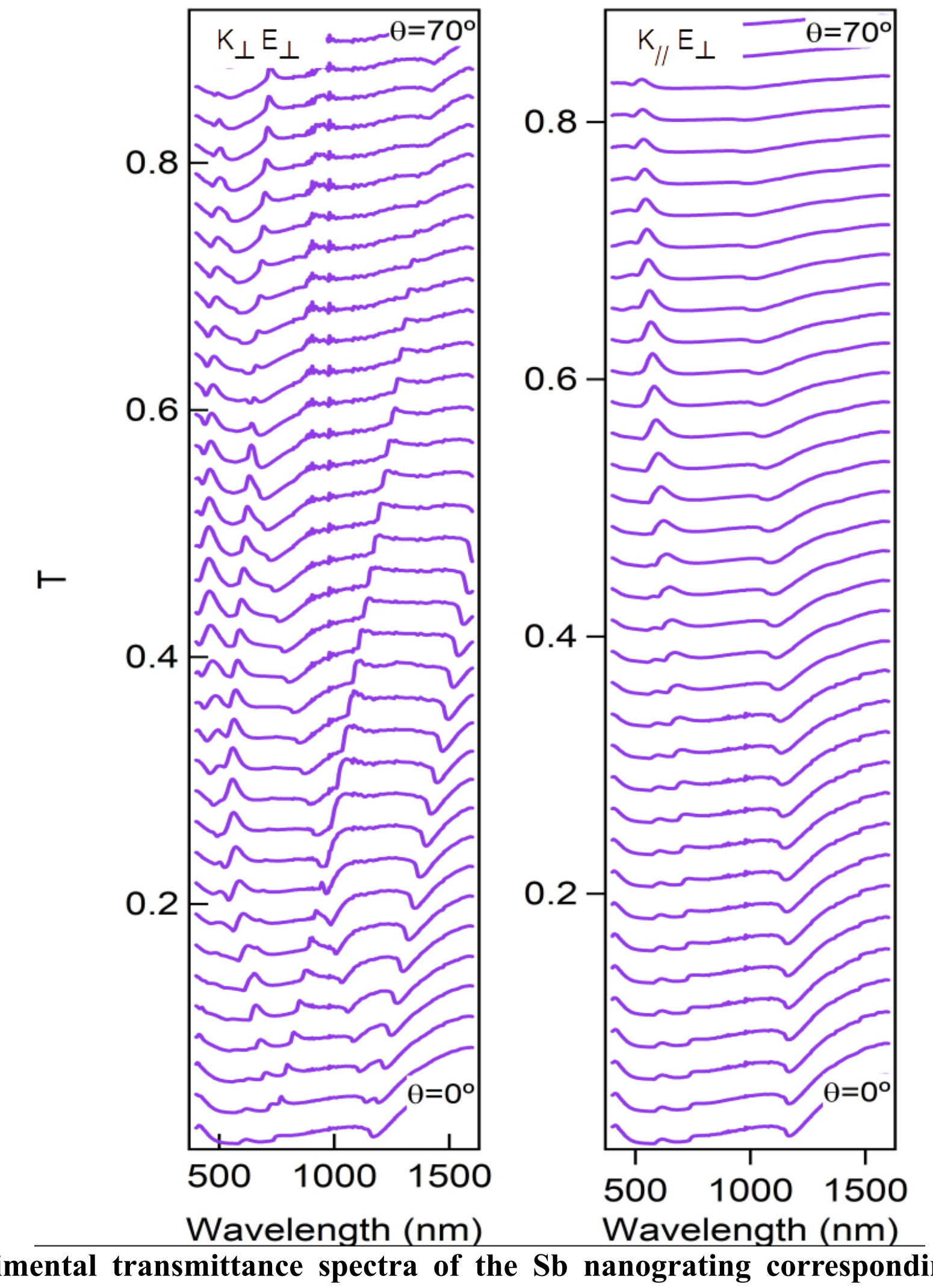

**Figure S5.1. Experimental transmittance spectra of the Sb nanograting corresponding to the color maps shown in Figure 3c.** The spectra are offset vertically with a 0.025 offset. The bottom and top spectra were measured at a 0º and 90º angles of incidence. The intermediate spectra were measured at increasing angles of incidence with a 2º step.

## S6. Sb nanograting: FDTD experiment vs simulation

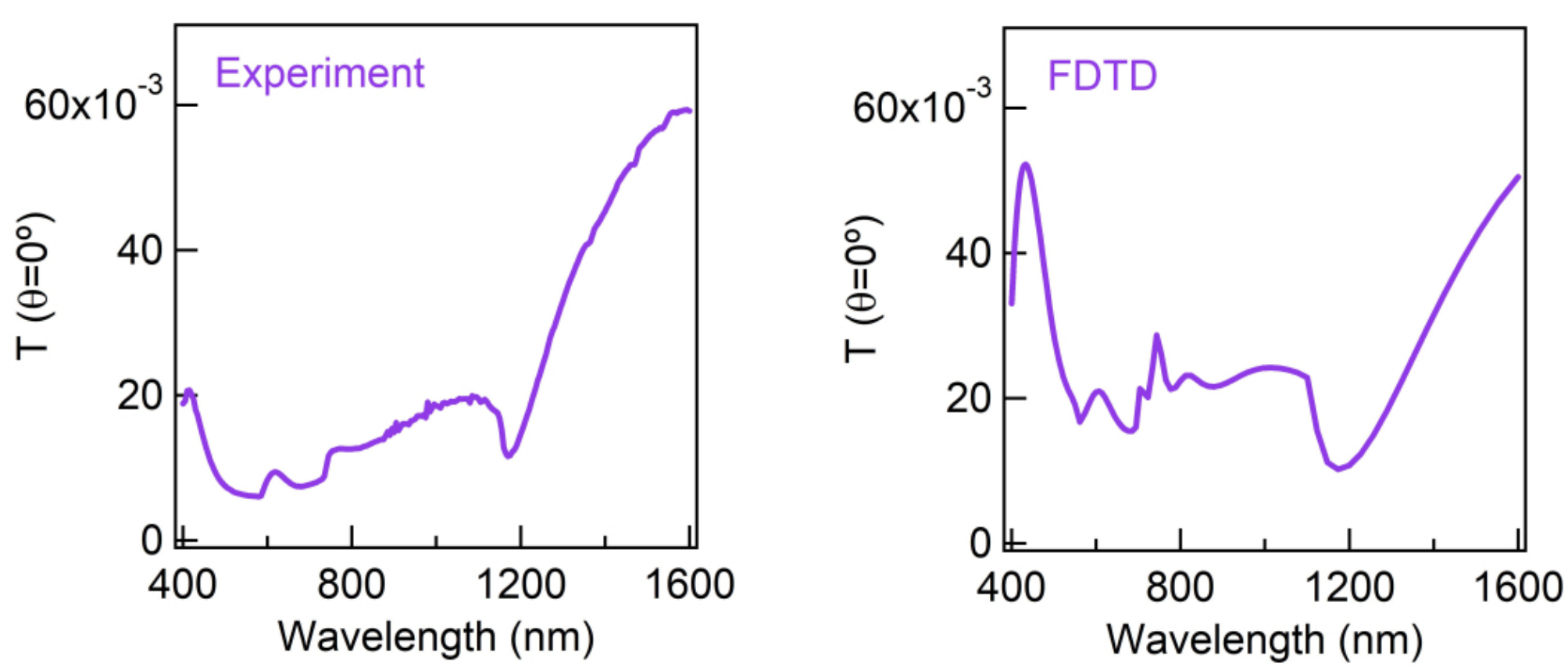

**Figure S6.1. Experimental and FDTD simulated transmittance spectrum of the Sb nanograting.** FDTD simulations yield a spectrum very similar to the experimental one. The experimental transmittance displays a downward trend upon decreasing the wavelength, which is not reproduced by FDTD simulations. This downward trend is due to light scattering by the rough nanograting surface, which is not taken into account in the FDTD simulations.

## S7. Sb nanograting: transmittance for the incident electric field oriented parallel to the lines

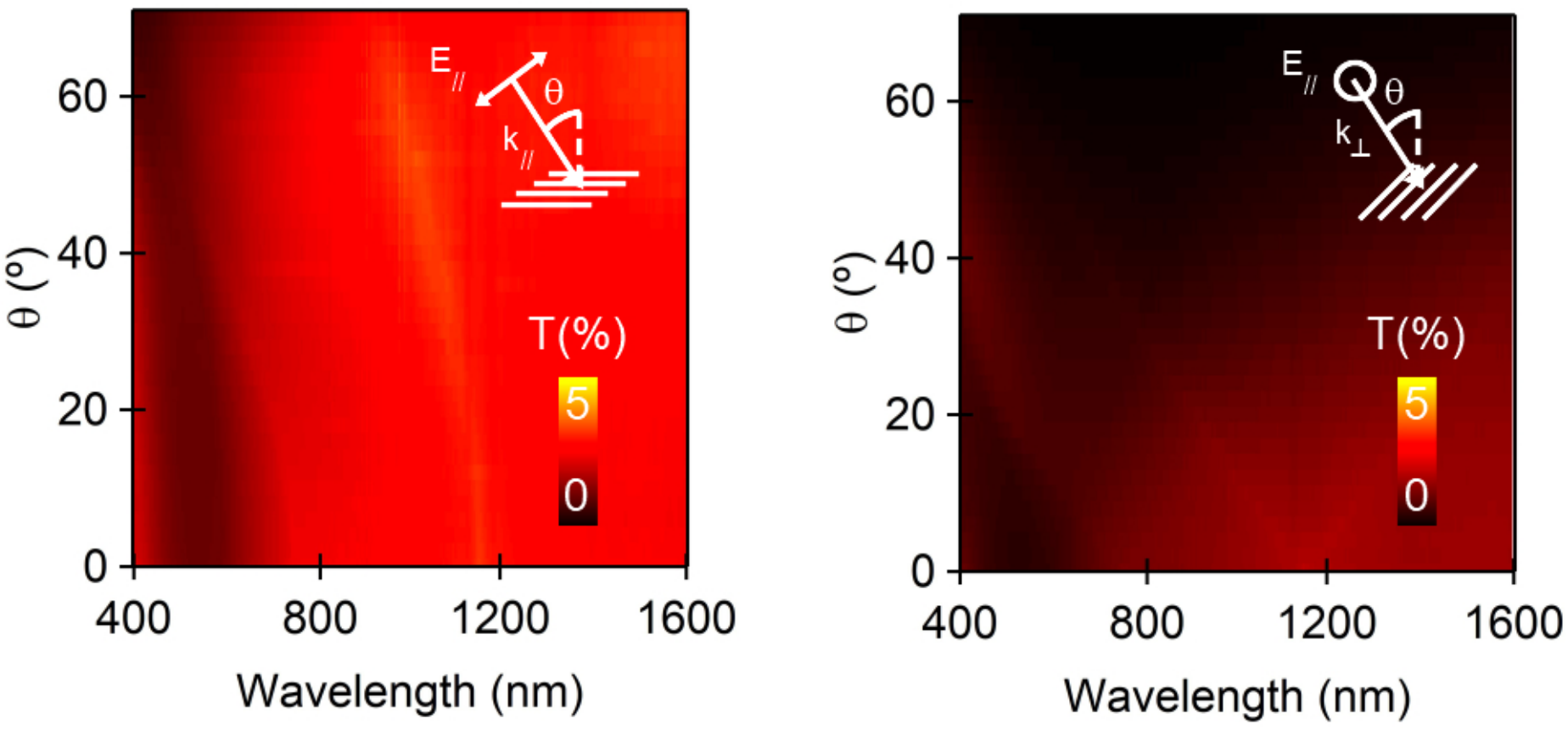

**Figure S7.1. Color maps of the experimental transmittance T of the Sb nanograting, complementary to those shown color maps shown in Figure 3c.** They were measured for the incident electric field parallel to the lines (E//) and with the incident wavevector parallel or perpendicular to the lines ($k_{//}$ or $k_{\perp}$). Transmittance is represented as a function of wavelength and the angle of incidence θ.

## S8. Sb nanograting: bulk refractive index sensing properties

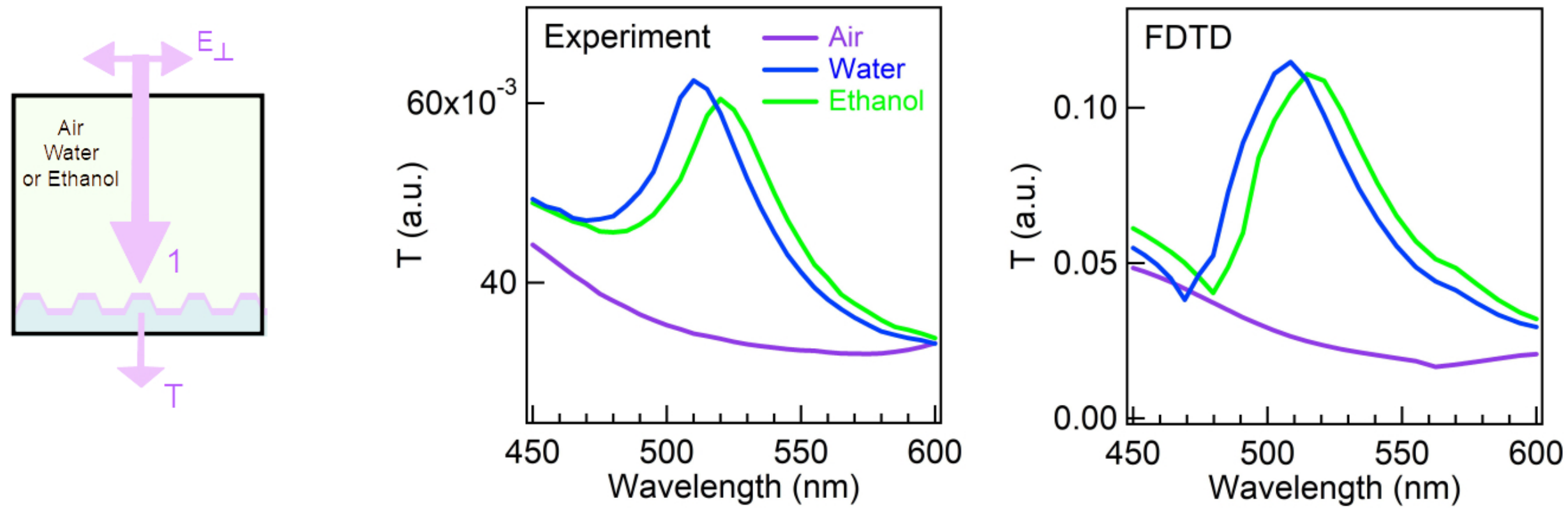

**Figure S8.1. Proof-of-concept of refractive index sensing with the Sb nanograting.** The nanograting was placed inside a cuvette, which was empty ($n = 1$) or filled with water ($n = 1.33$) or ethanol ($n = 1.36$). Its optical transmittance was measured at normal incidence. A resonance appeared in the visible region when filling the cuvette and shifted when the liquid was changed from water to ethanol. The same trend was observed by FDTD measurements considering that the nanograting is immersed in semi-infinite media with different refractive index ($n = 1$, $n = 1.33$, $n = 1.36$). This provides a proof of concept of bulk refractive index sensing with the Sb nanograting.